# Measurement of two independent phase-shifts using coupled parametric amplifiers

**Utsab Khadka,[1] Jiteng Sheng,[1] Xihua Yang [1,3] and Min Xiao[1,2]**
[1]Department of Physics, University of Arkansas, Fayetteville, Arkansas 72701, USA
[2]National Laboratory of Solid State Microstructures and Department of Physics, Nanjing University, Nanjing 210093, China
[3]Department of Physics, Shanghai University, Shanghai 200444, People's Republic of China
E-mail: ukhadka@uark.edu and mxiao@uark.edu

**Abstract.** In this article, we demonstrate a scheme capable of two-phase measurement, i.e. the simultaneous measurement of the two phase-shifts occurring in two independent Mach-Zehnder interferometers using one intensity detector. Our scheme utilizes dark-state-enhanced coupled parametric amplifiers in an atomic medium to mix the multiple fields probing the various arms of the interferometers in parallel. The two phase-differences are then encoded in separate continuous-variable parameters in the spectral waveform of the parametrically amplified atom-radiated signal field, which can be directly decoupled in a single intensity measurement. Besides resolving two phase differences in parallel, this method can also be used to increase the channel capacity in optical and quantum communication by the simultaneous use of phase-modulation and amplitude-modulation.

**Contents**

## 1. Introduction

Interference between two optical fields has been ubiquitously used in metrology [1,2], including in the measurement of length. When the wavelengths of the two fields are identical, there are measurement schemes in which the identical wavelengths are static, as well as scanned with time. In the first scheme, the measurement is performed at a single position in frequency space. Changes in the optical path length in one arm will alter the output field intensity. Examples include the traditional Mach-Zehnder and Michelson interferometers, and some state-of-the-art applications using this scheme include the measurement of minute space dilations arising from general relativistic effects [3]. In this scheme, one can measure changes in the relative path length difference between the two arms of the interferometer, but not the absolute path length difference between them.

   In the second scheme, which is a variant of the Mach-Zehnder interferometer, the wavelength of the two fields probing the interferometer's arms is scanned in time, and measurement is performed along a spectral line [4]. The reference arm's length is made different from the test arm's length, so that the phase-difference between the two beams evolves linearly along the spectral range being scanned. As a result, the output intensity of the interferometer will consist of fringes in frequency space. Here, changes in the path length of the test arm will alter the phase of the fringes. In addition to the fringe phase, which measures changes in the relative path length difference between the two arms of the interferometer, the fringe period (i.e. the spectral separation between two fringe maxima) measures the absolute path length difference between these two arms. This scheme has been popularly labeled "absolute distance interferometry" and has been utilized, for instance, in ATLAS, the largest particle detector of the Large Hadron Collider (LHC) project at CERN [4].

In the schemes described above, the optical fields undergo only linear transformations, namely that of propagation along the interferometer's arms, and transmission, reflection and additive mixing at beam splitters, before being measured by an intensity detector. There also exist interferometric schemes that use nonlinearities and multiplicative wave-mixing elements in order to process the field phases. Nonlinear wave-mixing processes such as two-photon absorption [5] and four-wave mixing [6-9] as well as closed-loop atomic interferometers [10-12] have been considered, and features such as the interference between multiple quantum transition amplitudes, squeezing and the parametric amplification within the interferometer have been utilized for enhancing sensitivity, resolution and visibility. Quantum interferometry is an active field of research, and exotic states of light are being tested as interferometric probes [13-15].

All of the schemes described above involve a single-phase measurement; that is, from the output field, one can extract information about changes occurring in the phase difference between one pair of optical fields. To our knowledge, there is no interferometer that can measure more than one phase difference in a single measurement. By a single measurement, we mean a measurement performed in a single spatial window (i.e. one detector) within a single temporal window (i.e. simultaneously).

In this article, we demonstrate a novel scheme showing the possibility of two-phase interferometry. The scheme is capable of processing the phase-differences of two independent pairs of optical fields in parallel and encoding them in separate continuous-variable parameters (phase and brightness) of a single output signal field that can be directly decoupled in a single intensity measurement. The key lies in merging ideas from both of the schemes described above; i.e. we shape the output signal's spectral bandwidth into fringes so that the fringe phase measures changes in one interferometer, while the fringe brightness measures changes in a second interferometer. The use of such capacity is twofold. First, it can be used to measure the phase-difference information of multiple interferometers in a single measurement. Second, it can be used to generate signals with increased phase-sensitive information encoded per channel.

In order to encode four optical field phases (i.e. two phase differences) in the intensity of one output field in a readily distinguishable way, we mix the fields in a nonlinear medium capable of effective phase-sensitive parametric amplification. We couple the two field pairs probing the two interferometers to two coexisting and coupled quantum nonlinear pathways in an atomic medium, the details of which will be described in the experimental section below. The individual amplifiers, which act as multiplicative wave mixers, are additively coupled via identical phase-matching so that their relative phase alters the resultant signal intensity even without using an external local oscillator; this relative phase is sensitive to one interferometer. The second interferometer causes identical fringes in both of the amplifier's responses, and its phase shift is measured by a spectral translation of these fringes. Dark-state resonances [16-22] are included in the parametric amplifiers in order to resonantly enhance the multi-photon transition amplitudes with suppressed losses for both driving and generated fields [23-33], as well as to attain low background noise and high resolution. We will first describe the experimental scheme and derive the equations, after which we will discuss the observations and results.

## 2. Experimental Setup

Our experimental scheme is shown in figure 1. We coherently drive the third- and fifth-order nonlinearities in an inverted-Y energy level configuration in rubidium atomic vapour, which is magnetically shielded and heated to 75° C. The driven nonlinearities are coupled by sharing common atomic transitions and driving beams ($\mathbf{E_i}$, $\mathbf{E_i}'$; i = 1, 2, 3) to radiate four-wave mixing (FWM) and six-wave mixing (SWM) signals in the same phase-matched mode (direction $\mathbf{k_m}$, frequency $\omega_m$). An avalanche photo-diode (APD) placed in the phase-matched direction measures the intensity of the resultant multi-wave-mixing signal. The pair of driving beams $E_i = e^{-i(\omega_i t - k_i z + \Phi_i)}$ and $E_i' = e^{-i(\omega_i t - k_i' z + \Phi_i')}$ originate from the same narrow-linewidth continuous wave (cw) laser source $LS_i$ (i =1, 2, 3), and are thus phase-coherent. The beam frequencies $\omega_2$ and $\omega_3$ are held fixed at the atomic transition frequencies $\omega_{cb}$ and $\omega_{db}$, respectively, while the frequency $\omega_1$ is swept across the Doppler-broadened spectral bandwidth of the $\omega_{ba}$ transition (where $\omega_{jk} = (E_k - E_j)/\hbar$ with $E_k$ the energy of the atomic level |k>). The weak probe beam $E_p = E_1 + E_1'$, travelling along the **z** direction, counter-propagates with the rest of the driving beams at small angles. Beam $E_2$ travels along the **–z** direction. At any plane transverse to the probe beam's direction, the driving beams pass through the four corners of a square with $E_3'$ furthest to $E_2$. Each side of the square subtends an angle of 0.35° at the center of the vapor cell, where all beams intersect. At the intersection region, the powers and diameters of the Gaussian beams $E_1$, $E_1'$, $E_2$, $E_2'$, $E_3$, $E_3'$ are approximately 3 mW, 3 mW, 30 mW, 4 mW, 65 mW, 65 mW and 0.5 mm, 0.6 mm, 1.3 mm, 1.4 mm, 0.7 mm, 0.6 mm, respectively.

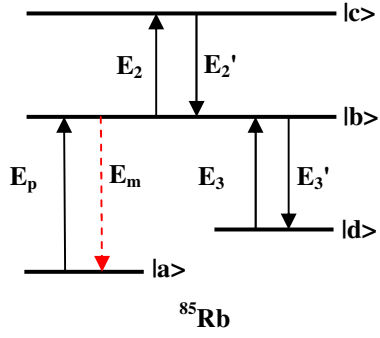

(a) Atomic configuration

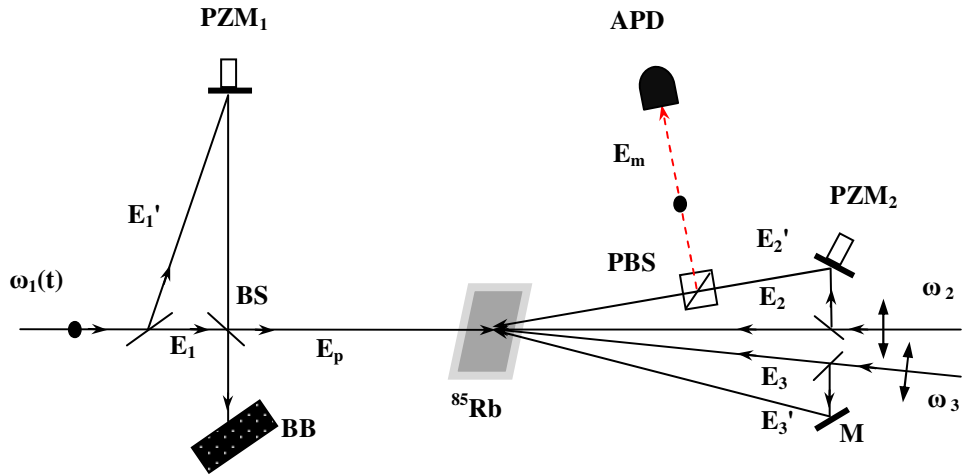

(b) Experimental setup

**Figure 1**. **Atomic configuration and experimental setup.** (a) Here, |a> and |d> are the hyperfine levels 2 and 3 of the $5S_{1/2}$ ground state, respectively; |b> and |c> correspond to the excited states $5P_{3/2}$ and $5D_{5/2}$, respectively. (b) (BS = 50/50 beam splitter, PBS = polarization beam splitter cube, BB = beam block, PZM = piezo-actuated mirror, M = rigid mirror, APD = avalanche photodiode). Here all the beams are shown in the same plane to visually "unfold" the MZ interferometers. In the actual setup, $E_2'$ and $E_3'$ lie in a plane that crosses the plane containing $E_p$, $E_2$ and $E_3$ inside the rubidium vapour cell (see text).

Due to this counter-propagating beam geometry, we obtain a Doppler-free two-photon-resonant (TPR) electromagnetically-induced-transparency (EIT) [16-22] coherence for the ladder-type subsystem (TPR frequency = $\omega_1 + \omega_2$) but not for the lambda-type subsystem (TPR frequency = $\omega_1 - \omega_3$). As will be shown below, this ladder-type EIT coherence enhances the FWM signal $E_f$ (phase-matching wave-vector $k_f = k_1 + k_2 - k_2' \equiv k_m$ and frequency $\omega_f = \omega_1 + \omega_2 - \omega_2' = \omega_1$) and SWM signal $E_s$ (phase-matching wave-vector $k_s = k_1 + k_2 - k_2 + k_3 - k_3' \equiv k_m$ and frequency $\omega_s = \omega_1 + \omega_2 - \omega_2 + \omega_3 - \omega_3' = \omega_1$). At the line-center of the Doppler- broadened transition from |a> to |b>, due to a large ground-state population, only these EIT-supported signals experience negligible absorption [22, 31-33], and all other signal fields have a vanishing transmission. In most of what follows, we limit our treatment to these two signal fields which spectrally coexist at the line center, denoted by $E_c = E_f + E_s$.

Before interacting with the atomic medium, the driving beams are made to probe three Mach-Zehnder interferometers $MZ_1$, $MZ_2$, and $MZ_3$. The optical path length difference $\Delta L_i$ between the two arms of $MZ_i$, and the resulting phase difference $\Delta\Phi_i$, is probed by the pair of beams $E_i$, $E_i'$ (i = 1, 2, 3). $\Delta L_3$ is held fixed, whereas $\Delta\Phi_1$ and $\Delta\Phi_2$ are the variables to be measured, or alternately, the parameters that can be controllably designed to phase-modulate the phase-matched signals resulting in $E_c$. Here, we discuss the measurement process.

In order to measure the two phases $\Delta\Phi_1$ and $\Delta\Phi_2$ in a single spatial-temporal reading of the signal field's intensity $|E_c|^2$, we decouple the effects of the two phases to two different continuous-variable (CV) parameters in the spectral waveform of the measurable signal intensity: the phase and amplitude of the intensity fringes, respectively. The fringes are generated in the spectral domain by modifying $MZ_1$ into a frequency-swept interferometer with unbalanced arms. The resultant probe beam then becomes $E_p = \frac{E_1 + E_1'}{\sqrt{2}} = \frac{A_1}{\sqrt{2}} e^{-i(\omega_1 t - k_1 z)}[1 + e^{-i\Delta\Phi_1(\omega_1, \Delta L_1)}]$, where $A_1 = A_1'$ are the field amplitudes and $\Delta\Phi_1(\omega_1, \Delta L_1) = \Gamma^{-1}\Delta\omega_1 + k_1 \Delta L_1$. That is, for a fixed but nonzero $\Delta L_1$, this setup causes the phase difference $\Delta\Phi_1$ to evolve linearly in the spectral domain as the probe beam frequency $\omega_1$ is swept across the atomic resonance linewidth. We have defined $\Gamma$ (2πHz) = c $\Delta L_1^{-1}$ to be the spectral period in which $\Delta\Phi_1$ evolves by 2π. When $\Delta L_1 \approx 10^7 \lambda_1$, a small change in the position of the mirror $PZ_1$ (typically a fraction of $\lambda_1$) has a negligible effect on $\Gamma$, and basically modulates only the second term of $\Delta\Phi_1(\omega_1, \Delta L_1)$.

As will be shown below, this phase-information encoded in the output of $MZ_1$ propagates through several orders of quantum nonlinear pathways in the phase-coherently driven medium, and is reproduced in both the FWM and SWM signals amplified by the coupled $\chi^{(3)}$ and $\chi^{(5)}$ processes. First, we focus on the FWM signal field $E_f = \eta_f \chi^{(3)} (E_2)^* E_2 E_p$, where the product of the FWM efficiency $\eta_f$ and third-order susceptibility comprises the frequency detuning factors, relaxation rates, dipole moment strengths, atomic density, and beam Rabi frequencies. The phase differences in the two interferometers $MZ_1$ and $MZ_2$ are encoded in the field envelope of this signal: $E_f = A_f e^{-i(\omega_f t - k_f z)} e^{-i\Delta\phi_2(\Delta L_2)}[1 + e^{-i\Delta\Phi_1(\omega_1, \Delta L_1)}]$, where $A_f = \eta_f \chi^{(3)} \frac{A_2' A_2 A_1}{\sqrt{2}}$ is a real amplitude. Note that we have replaced $(A_2')^*$ by $A_2'$, as the amplitude of the beam is held fixed and does not oscillate, and we also assume no depletion for the strong driving beam. Next, in the SWM channel, which utilizes the same EIT window supporting the FWM process, the field $E_2'$ is blocked. Instead, the field $E_2$ is used twice, and the SWM pathway is completed by using $E_3$ and $E_3'$ to drive transitions between the energy levels |b> and |d>. The SWM signal field is $E_s = \eta_s \chi^{(5)} E_3 (E_3')^* (E_2)^* E_2 E_p = A_s e^{-i(\omega_s t - k_s z)} e^{-i\Delta\phi_3(\Delta L_3)}[1 + e^{-i\Delta\Phi_1(\omega_1, \Delta L_1)}]$, where $A_s = \eta_s \chi^{(5)} \frac{A_3 A_3'(A_2)^2 A_1}{\sqrt{2}}$ is a real amplitude. The phase of $E_2$ has no contribution to the field in this pathway.

## 3. Results

When $E_f$ (or $E_s$) is observed individually, the phase $\Delta\Phi_2(\Delta L_2)$ (or $\Delta\Phi_3(\Delta L_3)$) arising from $MZ_2$ (or $MZ_3$) does not have an observable effect on the intensity. However, it is obvious that the phase $\Delta\Phi_1(\Delta L_1, \omega_1)$ arising from $MZ_1$ will cause identical oscillations in the intensities of the FWM and SWM signals across their spectral bandwidths, even without the use of an additional LO. When the two EIT-coupled spectrally coexisting signals are also phase-matched ($k_f = k_s = k_m$) and polarization-matched, all of which are achieved by our specially-designed beam geometry, they interfere. Since $E_2'$ contributes only to $E_f$, we can tune the strength of this driving field to attain identical strengths for $E_f$ and $E_s$. The resultant amplified signal field at the line center becomes $\mathbf{E_c} = \mathbf{E_f} + \mathbf{E_s} = \mathbf{A_c}\{e^{-i\Delta\phi_2(\Delta L_2)}[1 + e^{-i\Delta\Phi_1(\omega_1, \Delta L_1)}] + e^{-i\Delta\phi_3(\Delta L_3)}[1 + e^{-i\Delta\Phi_1(\omega_1, \Delta L_1)}]\}$, where $\mathbf{A_c}$ is the complex field $\mathbf{A_c} = A_c e^{-i(\omega_1 t - k_m z)}$ with real amplitude $A_c = A_f = A_s$. The path length difference $\Delta L_3$ in $MZ_3$ is held fixed at $\Delta\Phi_3 = 0$, thereby reducing the signal's dependence to the two variable phases $\Delta\Phi_1$ and $\Delta\Phi_2$. The resulting signal intensity within the EIT-supported spectral bandwidth at the line center would thus be

$$I_c = (I_{c0}/4)\{1 + \cos[\Delta\Phi_1(\omega_1, \Delta L_1)]\}\{1 + \cos[\Delta\Phi_2(\Delta L_2)]\} \tag{1(a)}$$

$$= (I_{c0}/4)\{1 + \cos[\Gamma^{-1}\Delta\omega_1 + k_1 * \Delta L_1]\}\{1 + \cos[k_2 * \Delta L_2]\} \tag{1(b)}$$

where $I_{c0}$ is the maximum fringe brightness (amplitude) occurring at $\Delta\Phi_1 = 0$ and $\Delta\Phi_2 = 0$. The linear dependence of $\Delta\Phi_1$ with $\omega_1$ causes intensity fringes across the spectral bandwidth of the signal with spectral period $\Gamma$ ($2\pi$ Hz). Any additional phase change in $\Delta\Phi_1$, for instance due to a small shift in $PZ_1$ or any other phase shifting element placed in $MZ_1$, will result in a translation of the intensity fringes in the spectral domain spanned by $\omega_1$ (figure 2(a)). The phase change can then be inferred from the spectral displacement, i.e. the spectral-domain phase shift, of the fringes. A change in $\Delta\Phi_2$, on the other hand, would alter the brightness of each fringe without altering their spectral positions (figure 2(b)). Equation 1 thus shows the main result of this letter, namely the resolvable coupling of two different phase differences to two different continuous-variable parameters of a single intensity measurement.

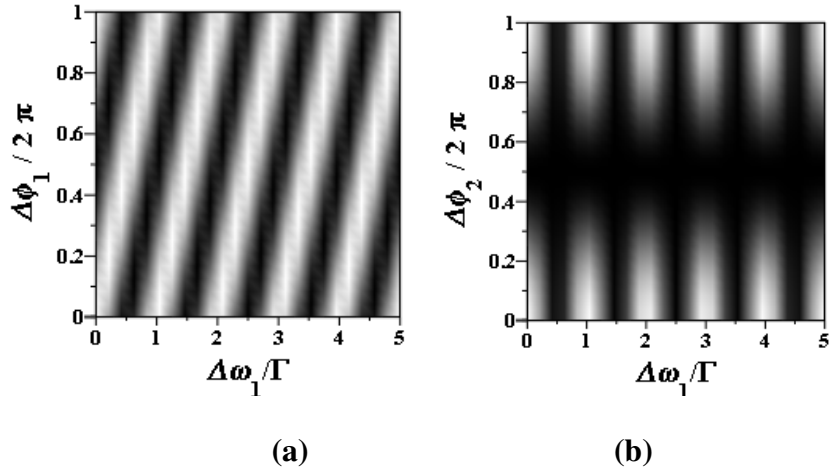

**(a)**          **(b)**

**Figure 2. Theoretical plots for the signal intensity $I_c$. (a)** Three-dimensional space spanned by $I_c$, $\Delta\omega_1/\Gamma$ and $\Delta\phi_1/2\pi$ with $\Delta\phi_2$ held fixed **(b)** Three-dimensional space spanned by $I_c$, $\Delta\omega_1/\Gamma$ and $\Delta\phi_2/2\pi$ with $\Delta\phi_1$ held fixed. In the grayscale intensity representation of $I_c$, white = bright fringe and black = dark fringe. Any phase-difference modulation in the interferometer $MZ_1$ ($MZ_2$) is measurable as a phase-shift (amplitude-modulation) of the intensity fringes occurring when frequency is scanned.

For the beam geometry being considered, the Doppler-broadened FWM signal $E_D$ driven by $E_p$, $E_3$ and $E_3'$ in the lambda-type subsystem also travels along $k_m$ ($k_D = k_1 + k_3 - k_3' \equiv k_m$). However, since it is not supported by EIT, it is completely absorbed at the line center, and occurs only at the wings of this Doppler-broadened transition. That is, $E_D$ is spectrally isolated, and does not coexist with $E_c = E_f + E_s$. Since $E_D$ is proportional to the product $E_3(E_3')^* E_p$, it is also affected by the modulation in $\Delta\Phi_1$. Due to the large spectral bandwidth of this signal, it might be useful to utilize it in conjunction with the EIT-bandwidth-limited signal $E_c$ for measuring $\Delta\Phi_1$. However, unlike $I_c$, this signal's intensity $I_D$ does not contain the information of two phases, which is the primary objective of this article.

Figures 3 and 4 show the photocurrents measured by the APD for various frequency detunings and phases, when $\Delta L_1 = 7.84$ m ($\Gamma = 2\pi \times 38$ MHz). In figure 3((i)), the left box corresponds to the spectral bandwidth supported by the ladder-type EIT coherence, and shows the resultant intensity of the coexisting Doppler-free signals, $I_c$. In this spectral bandwidth, occurring at the line-center of the |a> → |b> transition, the other signals that are not EIT-supported vanish. The right box corresponds to the spectral region towards the blue-detuned wing of the Doppler-broadened transition, where the spectrally broad FWM signal $E_D$ becomes measurable due to reduced absorption. The phase modulation in $MZ_1$ is evident in all cases. By keeping all other experimental parameters identical but shifting $PZ_1$ to alter $\Delta L_1$ by $\lambda_1/2$, creating a $\pi$ phase-shift between the $MZ_1$ beams, we observe a spectral translation of the fringes by $\Gamma/2$ (figure 3(ii)), while the amplitude of the peaks and envelope remain fixed. When $\Delta\Phi_2$ is altered, the spectral positions of the intensity fringes remain unchanged. However, in the spectral region containing the two coexisting signals, $\Delta\Phi_2$ modulates the amplitude of the fringes. Figure 4 shows $I_c$ for three different values of $PZ_2$, corresponding to variations in the $MZ_2$'s phase $\Delta\Phi_2$.

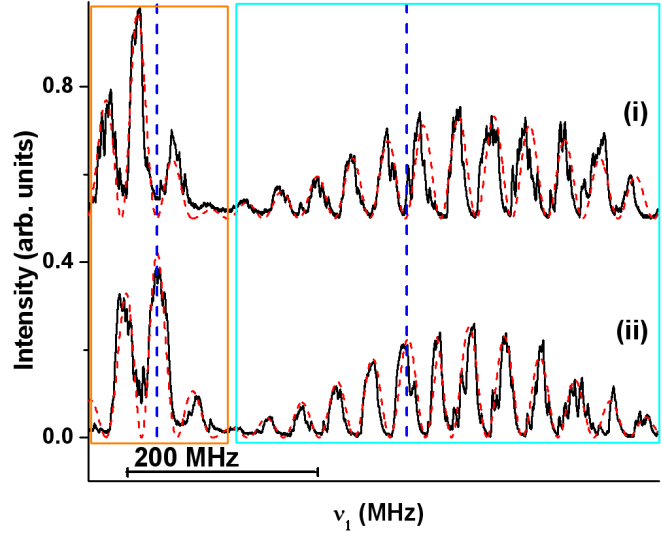

**Figure 3. Observations for a varying $\Delta\Phi_1$.** Experimental traces (black, solid) and theoretical fits (red, dashed) of the parametrically amplified signals. The left and right boxes highlight the spectral regions that amplify $E_c$ and $E_D$ (corresponding to the center and the blue-detuned regions of the Doppler-broadened D2 transition), respectively. When $\Delta\Phi_1$ is increased by $\pi$, the fringes in the upper trace **(i)** spectrally translate by $\Gamma/2$ in the lower trace **(ii)**. The two blue vertical dashed lines are visual guides for two spectrally fixed positions. All other parameters, including $\Delta\Phi_2$, are held fixed.

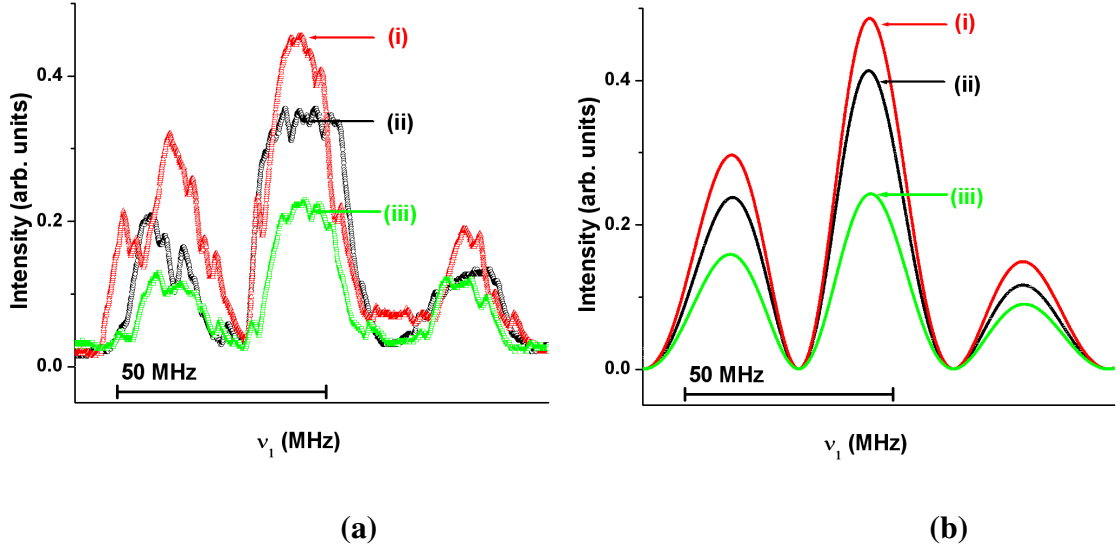

**Figure 4. Observations for a varying $\Delta\Phi_2$.** Experimental traces **(a)** and theoretical plots **(b)**, showing $I_c$ for three different values of $\Delta\Phi_2$: (i) 0 (ii) $\pi/3$ (iii) $\pi/2$. All other parameters, including $\Delta\Phi_1$, are held fixed. Here, it is the brightness of the fringes that changes. The spectral region corresponds to the left box shown in Fig. 3.

## 4. Conclusions and Outlook

In conclusion, we have demonstrated a scheme for measuring the phase differences in two different Mach-Zehnder interferometers $MZ_1$ and $MZ_2$ in a single measurement of the signal intensity. The key lies in using a spectrally broad measurement in order to have two continuous-variable observables in the intensity: the phase and the brightness of the spectral-domain fringes. The two phases to be measured are then coupled to the two observables, respectively. To our knowledge, this is the first demonstration of identifying relative changes in two pairs of optical path lengths in a single intensity measurement, and might be useful in increasing the spatial dimensions being probed in interferometric measurements, for instance in the Laser Interferometer Gravitational-Wave Observatory (LIGO) [3]. Having phase-sensitive control over two continuous-variable intensity parameters also increases the information capacity per channel [34], which might be useful in optical and quantum communication. In particular, if one set of information is carried by the amplitude of the fringes (via amplitude-modulation), a second set of information can now be simultaneously and separately encoded in the phase of the fringes (via phase-modulation). Another advantage of the scheme is that an external local oscillator (LO) is not needed while measuring the multiple phases. The coexisting fields that are parametrically amplified in the phase-matched mode sufficiently produce the necessary interference and intensity variations at the detector. The lack of need for a LO could make the method valuable in multi-party, long-distance communication of phase-modulated signals.

**Acknowledgements**

Partial funding support from the National Science Foundation is acknowledged. MX acknowledges partial support from CNSF (#11021403).